# P-V-T Equation of State for Periclase

Jozsef Garai[a,*], Jiuhua Chen[a, b], and Gabor Telekes[c]

[a]*Department of Mechanical and Materials Engineering, Florida International University, Miami, USA*
[b]*CeSMEC, Florida International University, Miami, USA*
[c] *Ybl Miklós Faculty of Architecture and Civil Engineering, St. István University, Budapest, Hungary*

A B S T R A C T

Collecting the complete data set of previous experiments on periclase, covering a pressure and temperature range of 0-141.8 GPa and 100-3031 K respectively, the first comprehensive P-V-T description of MgO is presented comprising all previous experiments. The P-V-T EoS of Birch-Murnaghan, Rydberg-Vinet and Garai are determined by unrestricted fitting. The three EoSs are consistent and a unique set of parameters is able to cover the entire pressure and temperature range. The RMS misfits for the pressure are 0.371 GPa, 0.381 GPa and 0.396 GPa for the Garai, Birch-Murnaghan and Rydberg-Vinet EoSs. The RMS misfits for the volume and the temperature are 0.018 $cm^3$ and 60.3 K for the EoS of Garai.

## 1. Introduction

MgO has low chemical reactivity and it is stable in large pressure and temperature range which makes it an ideal pressure calibrant. Periclase is the end member of the (Mg,Fe)O solid solution series. Mg-rich ferropericlase is the second most abundant component of the Earth's lower mantle [1, 2] and has significant interest in geophysics. Precise description of the pressure, volume and temperature () relationship of MgO is therefore essential.

The description of solids can be given in various ways like implying pressure increase first, followed by heating at constant volume or at constant pressure, etc. The most common approach is to raise the temperature first and determine the parameters of the isothermal EoS for the given temperature and then by using the isothermal EoSs the effect of pressure is calculated [e.g. 3].

The most widely used isothermal EoSs are the Birch-Murnaghan (B-M) and the Rydberg-Vinet (R-V) EoSs. The third-order (B-M) EoS [4-6] is given as:

$$P=\frac{3K_{0T}}{2}\left[\left(\frac{V_{0T}}{V}\right)^{\frac{7}{3}}-\left(\frac{V_{0T}}{V}\right)^{\frac{5}{3}}\right]\left\{1+\frac{3}{4}\left(K_0^{'}-4\right)\left[\left(\frac{V_{0T}}{V}\right)^{\frac{2}{3}}-1\right]\right\}, \tag{1}$$

where $K_{0T}$, $K_0^{'}$ and $V_{0T}$ are the isothermal bulk modulus, the pressure derivative of the bulk modulus and the volume respectively at zero pressure and at the temperature of interest.

The R-V EoS derived from a general inter-atomic potential [7]. This EoS was rediscovered by Rose et al., [8] and Vinet [9]. The EoS is given as:

$$P=3K_{0T}\frac{1-f_V}{f_V^2}e^{\left[\frac{3}{2}\left(K_0^{'}-1\right)\left(1-f_V\right)\right]} \tag{2}$$

where

$$f_V=\left(\frac{V}{V_{0T}}\right)^{\frac{1}{3}}. \tag{3}$$

The EoS is attributed to Vinet in most of the geophysical literature [10]. In order to give credit to Rydberg the EoS is called Rydberg-Vinet EoS here.

The parameters are usually available at ambient condition. Assuming linear temperature dependence, the values of $K_{0T}$ and $V_{0T}$ can be calculated for higher temperatures as:

$$K_{0T}(T)=K_{0T}(T_0)+\left(\frac{\partial K_{0T}}{\partial T}\right)_p(T-T_0). \tag{4}$$

and

$$V_{0T}(T)=V_{0T}(T_0)e^{\alpha_0(T-T_0)+\alpha_1\left(T^2-T_0^2\right)}. \tag{5}$$

Substituting Eqs. (4) and (5) into the isothermal EoSs [Eqs. (1)-(3)] results in description. The parameters of the EoS are inter-related and the use of confidence ellipses is suggested to determine the optimum values [e.g. 11].

## 2. Equation of States used in this investigations

In order to provide a comprehensive description of solids the effect of temperature has to be incorporated into the original isothermal EoSs. To make the calculations simple the ambient reference frame is replaced with an absolute reference frame. The word initial refers to thermodynamic parameter or quantity which is given at zero pressure and temperature. The initial volume $[V_o]$ is defined then as:

$$V_o \equiv nV_o^m \tag{6}$$

where n is the number of moles and $V_o^m$ is the initial molar volume at zero pressure and temperature. The initial bulk modulus is defined as:

$$K_o \equiv \lim_{p \Downarrow 0} K_{T=0} = \lim_{p \Downarrow 0} \left[ -V_{T=0} \left( \frac{\partial P}{\partial V} \right)_{T=0} \right]. \qquad (7)$$

The linear temperature dependence of the volume coefficient of thermal expansion [Eq. (5)] is not valid at temperatures lower than the Debye temperature [12] and $\alpha_o$ has zero value at zero temperature. However, the error introduced by the linear approximation is insignificant at temperatures higher than room temperature [13]. The volume at zero pressure is calculated then as:

$$V_{0T} = nV_o e^{(\alpha_0 + \alpha_1 T)T}, \qquad (8)$$

where $\alpha_o$ is the extrapolated value of the volume coefficient of thermal expansion at zero pressure and temperature.

Assuming constant value for the product of the isothermal bulk modulus and the volume coefficient of thermal expansion the temperature dependence of the bulk modulus at 1 bar pressure is derived from classical thermodynamic relationships [14]. The bulk modulus is given as:

$$K_{0T} = K_o e^{-\int_{T=0}^{T=T} \alpha_{V_p} \delta dT} \cong K_o e^{-(\alpha_0 + \alpha_1 T)\delta T} \qquad (9)$$

where $\delta$ is the Anderson- Grüneisen parameter, which defined as:

$$\delta \equiv \left( \frac{\partial \ln K_T}{\partial \ln V} \right)_p = -\frac{1}{\alpha_{V_p}} \left( \frac{\partial \ln K_T}{\partial T} \right)_p = -\frac{1}{\alpha_{V_p} K_T} \left( \frac{\partial K_T}{\partial T} \right)_p. \qquad (10)$$

The pressure effect on the bulk modulus is calculated as:

$$K_{p,T} = \left( K_o + K_o' p \right) e^{-(\alpha_0 + \alpha_1 T)\delta T}. \qquad (11)$$

Incorporating Eqs. (6)-(9) into the original isothermal EoSs [Eqs. (1)-(3)] results in P(V,T) EoSs. The P(V,T) form of Birch-Murnaghan EoS can be written as:

$$P = \frac{3K_o e^{-(\alpha_0 + \alpha_1 T)\delta T}}{2} \left[ \left( \frac{V_o e^{(\alpha_0 + \alpha_1 T)T}}{V} \right)^{\frac{7}{3}} - \left( \frac{V_o e^{(\alpha_0 + \alpha_1 T)T}}{V} \right)^{\frac{5}{3}} \right] \left\{ 1 + \frac{3}{4} \left( K_o' - 4 \right) \left[ \left( \frac{V_o e^{(\alpha_0 + \alpha_1 T)T}}{V} \right)^{\frac{2}{3}} - 1 \right] \right\} \qquad (12)$$

and the P(V,T) form of the Rydberg-Vinet EoS is:

$$P = 3K_o e^{-(\alpha_o+\alpha_1 T)\delta T} \frac{1-\left(\dfrac{V}{V_o e^{(\alpha_o+\alpha_1 T)T}}\right)^{\frac{1}{3}}}{\left(\dfrac{V}{V_o e^{(\alpha o+\alpha_1 T)T}}\right)^{\frac{2}{3}}} \, e^{\left\{\frac{3}{2}\left(K_0'-1\right)\left[1-\left(\frac{V}{V_o\, e^{(\alpha_o+\alpha_1 T)T}}\right)^{\frac{1}{3}}\right]\right\}} . \qquad (13)$$

Recently an eight parameters semi empirical EoS has been proposed and tested against the experiments of $MgSiO_3$ perovskite [13] with positive result. The EoS is given as:

$$V = nV_o e^{\frac{-P}{K_o+K_{P1}P+K_{P2}P^2}+\left(\alpha_o+\alpha_{P1}P+\alpha_{P2}P^2\right)T+\left(1+\frac{\alpha_{P1}P+\alpha_{P2}P^2}{\alpha_o}\right)^a \alpha_{T1}T^2} \qquad (14)$$

where, $K_{P1}$ is a linear, $K_{P2}$ is a quadratic term for the pressure dependence of the bulk modulus, $\alpha_{P1}$ is a linear and $\alpha_{P2}$ is a quadratic term for the pressure dependence of the volume coefficient of thermal expansion, $\alpha_{T1}$ is a linear term for the temperature dependence of the volume coefficient of thermal expansion and a is constant characteristic of the substance. The theoretical explanations for (14) and the physics of the parameters are discussed in detail [13]. The equation has an analytical solution for the temperature

$$T = \frac{-\left(\alpha_o+\alpha_{P1}P+\alpha_{P2}P^2\right) \pm \sqrt{\left(\alpha_o+\alpha_{P1}P+\alpha_{P2}P^2\right)^2+4\alpha_{T1}\left(1+\dfrac{\alpha_{P1}P+\alpha_{P2}P^2}{\alpha_o}\right)^a \Bigg[\ln\left(\dfrac{V}{V_o}\right)+\dfrac{P}{K_o+K_{P1}P+K_{P2}P}}}{2\alpha_{T1}\left(1+\dfrac{\alpha_{P1}P+\alpha_{P2}P^2}{\alpha_o}\right)^a} \qquad (15)$$

The pressure can be determined by repeated substitutions as:

$$P = \lim_{n\to\infty} f^n(P) \qquad (16)$$

where

$$f^n(P) = \left(K_o+K_{P1}P_{n-1}+K_{P2}P_{n-1}^2\right)\Bigg[\left(\alpha_o+\alpha_{P1}P_{n-1}+\alpha_{P2}P_{n-1}^2\right)T+\Bigg(1+\frac{\alpha_{P1}P_{n-1}+\alpha_{P2}P_{n-1}^2}{\alpha_o} \qquad (17)$$

$n \in \mathbb{N}^*$ and $P_0 = 0$

The convergence of Equation (17) depends on pressure. For the maximum pressure used in this study (up to 140 GPa) n = 15 is sufficient. The maximum convergence error $[\varepsilon]$ for the investigated data set is 0.05 GPa where

$$\varepsilon \geq \left|f^{15}(P)-f^{14}(P)\right| . \qquad (18)$$

In this study, collecting the complete data set of previous experiments the parameters of the modified EoSs of (B-M) [Eq.(12)] and (R-V) [Eq.(13)] and the EoS (G) [Eq. (14)] are determined for periclase.

## 3. Data Analyzes

The fitting accuracy of the EoSs is evaluated by RMS misfits and Akaike Information Criteria (AIC) [15, 16]. The Akaike Information Criteria is devised assessing the right level of complexity. Assuming normally distributed errors criterion is calculated as:

$$AIC = 2k + N \ln\left(\frac{RSS}{N}\right), \quad (19)$$

where RSS is the residual sum of squares, k is the number of parameters and N is the number of observations or data. AIC penalizes both for increasing the number of parameters and for reducing the size of data. The preferred model is the one which has the smallest AIC value.

MgO has been subject to numerous experimental investigations [e.g. 17-33] and computational and theoretical studies [e.g., 34-42]. Previous studies [25, 30] demonstrated that the deviatoric stress has a significant effect on the lattice parameters; therefore, in this study we consider only experiments conducted under hydrostatic or semi-hydrostatic conditions [23, 25, 27, 28, 30, 32, 33, 40]. If two pressure scales are used to determine the pressure then the average of the two reported pressure is used. Thermal expansion [22, 26, 43], Brillouin scattering [44] rectangular parallelepiped resonance method [45-46] measurements at atmospheric pressure are also included in the data set.

## 4. Results

Parameters providing the best fit against the data are determined by unrestricted fitting for the three EoSs [Eqs. (12)-(14)]. Experiments with possible systematic errors (17) were dropped from the data set. The dropping in each cases were justified by AIC. The 406 experiments used in this study span the pressure and temperature range of 0-141.85 GPa and 100-3031 K respectively (Fig. 1).

The determined one set of parameters for each of the EoS is sufficient to cover the entire pressure and temperature range of the experiments. The parameters are reported in Table. 1.

The RMS misfits for the complete data set (N=406) are 0.371 GPa (G), 0.381 GPa (B-M), 0.396 GPa (R-V), and the AIC values are -804.5, -770.7 and -741.1 and respectively. The RMS misfit of the volume and temperature is 0.018 $cm^3$ and 60.3 K respectively for EoS (G). The residuals of the three EoSs are plotted on Fig. 2 and 3. The three EoS reproduces the data practically by the same uncertainty. The RMS misfits of the individual investigations are also calculated by using the parameters determined for the entire data set. The values are given in Table 3.

The parameters of the EoS allows calculating the volume coefficient of thermal expansion and the bulk modulus for a given pressure and temperature. The volume coefficient of thermal expansion from the EoS of B-M and R-V can be calculated as:

$$\alpha = \alpha_o + \alpha_1 T \qquad (20)$$

and from the EoS G as:

$$\alpha = \alpha_o + \alpha_{P1} P + \alpha_{P2} P^2 \left(1 + \frac{\alpha_{P1} P + \alpha_{P2} P^2}{\alpha_o}\right)^a \alpha_{T1} T . \qquad (21)$$

The bulk modulus can be calculated by using Eq. (9) for the EoSs of B-M and R-V. Using the definition of the bulk modulus

$$K_T \equiv -V \left(\frac{\partial P}{\partial V}\right)_T . \qquad (22)$$

and calculating the volumes using Eq. (14) the bulk modulus can be determined from the EoS G as:

$$K_{P,T} = \frac{-P}{\ln\left(\frac{V_{P,T}}{V_{P=0,T}}\right)} \qquad (23)$$

Equations (20)-(23) calculates the volume coefficient of thermal expansion and the bulk modulus for the entire temperature and pressure range. The temperature and pressure derivatives of the volume coefficient of thermal expansion and the bulk modulus are not zero for MgO. Thus equations (20)-(23) can not be used to calculate the parameters representing a specific pressure and temperature. The "instantaneous" value can be calculated by using the definitions of the volume coefficient of thermal expansion and the bulk modulus and substituting the volumes calculated by Eq. (14). The volume coefficient of thermal expansion of a specific pressure and temperature is given as:

$$\alpha = \lim_{\Delta T \Rightarrow 0} \frac{\ln\left[\frac{V\left(P, T + \frac{\Delta T}{2}\right)}{V\left(P, T - \frac{\Delta T}{2}\right)}\right]}{\Delta T}, \qquad (24)$$

and the "instantaneous" bulk modulus as:

$$K_{P,T} = \lim_{\Delta P \Rightarrow 0} \frac{-\Delta P}{\ln\left[\frac{V\left(P + \frac{\Delta P}{2}, T\right)}{V\left(P - \frac{\Delta P}{2}, T\right)}\right]}. \tag{25}$$

The calculated volume coefficient of thermal expansion values are fit well to experiments at ambient condition (Fig. 4). The fitting of the bulk modulus is not perfect (Fig. 5).

If the bulk modulus value is fixed to $K_{0T=298} = 161.6$ GPa (determined from high precision sound velocity measurements [46] and the fitting is repeated then the RMS misfit increases from 0.381 GPa (B-M) and 0.396 GPa (R-V) to 0.469 GPa (B-M) and 0.466 GPa (R-V). Fitting to higher temperature experimental values requires additional constrains on the volume coefficient of thermal expansion and the Anderson- Grüneisen parameter (Tab. 1). The RMS misfit then increases from 0.381 GPa (B-M) and 0.396 GPa (R-V) to 0.921 GPa (B-M) and 1.005 GPa (R-V). This high increase of misfit is also accompanied with a systematic error against high pressure experiments. Eventhough, the bulk modulus values do not fit well to experiments the EoS reproduce the experimental volumes with high accuracy at ambient conditions (Fig. 6).

The thermodynamic parameters of the EoSs are also tested against heat capacity measurements. The constant pressure molar heat capacity is calculated as:

$$c_p^s = c_{Debye} + \frac{VTB\alpha^2}{n}. \tag{26}$$

The Debye heat capacity is calculated by using the Debye function [47].

$$c_{Debye} = 3Rf \qquad f = 3\left(\frac{T}{T_D}\right)^3 \int_0^{x_D} \frac{x^4 e^x}{(e^x - 1)^2} dx \tag{27}$$

and

$$x = \frac{h\omega}{2\pi k_B T} \qquad \text{and} \qquad x_D = \frac{h\omega_D}{2\pi k_B T} = \frac{T}{T_D} \tag{28}$$

where h is the Planck's constant, $\omega$ is the frequency, and $\omega_D$ is the Debye frequency. Equation (27) has to be evaluated numerically [48]. In Eq. (26) the volume coefficient of thermal expansion calculated by Eq. (21) and the bulk modulus by Eq. (23) and (14). The calculated values agree well with the experimental data of 1 bar pressure [49, 50] (Fig. 7). The good fit to heat capacity experiments is an additional indication that the determined thermodynamic parameters of the EoSs are correct.

Using the set of parameters determined by unrestricted fitting the pressure corresponding to 300 K and 3000 K was calculated for all the three EoS. The differences between the calculated pressures are plotted on Fig. 8. The three EoSs are within 2.8 GPa and 3.5 GPa error margin in the entire pressure range at 300 K and 3000 K respectively. Up to 120 GPa the three EoSs are almost identical above this pressure small deviation occurs.

## 5. Conclusions

The entire data set of MgO has been collected. The 406 data are fitted against the modified $(\alpha K_T = \text{const.})$ Birch-Murnaghan, Rydberg-Vinet and Garai EoSs. All EoSs are able to reproduce the data with accuracy close the uncertainty of the experiments with one set of parameters for the entire pressure (0-141.8 GPa) and temperature (100-3031 K) range of the experiments. The uncertainties of the three EoSs are practically the same. The advantage of the EoS of Garai is that the volume and the temperature can be calculated directly.

**Acknowlegement**

The authors thank Sergio Speziale for reading and commenting the manuscript and Rostislav Hrubiak for the suggested bulk modulus calculation. The authors also thank the constructive criticism of the reviewers which helped to improve the quality of the manuscript. This research was supported by NSF Grant #0711321.

## References

[1] L. Stixrude, R.J. Hemley, Y. Fei and H-K Mao, Science, 257 (1992) 1099-1101.
[2] I. Jackson, Geophys. J. Int., 134 (1998) 291-311.
[3] T.S. Duffy, and Y. Wang, Reviews in Mineralogy, 37 (1998) 425–457.
[4] F. Birch, Phys. Rev. 71 (1947) 809-824.
[5] F.D. Murnaghan, Am. J. Math. 49 (1937) 235-260.
[6] F.D. Murnaghan, Proc. Nat. Acad. Sci., 30 (1944) 244-247.
[7] R. Rydberg, Z. Phys. 73 (1931) 376–385.
[8] J.H. Rose, J.R. Smith, F. Guinea, and J. Ferrante, Phys. Rev. B, 29 (1984) 2963-2969.
[9] P. Vinet, J.R. Smith, J. Ferrante, and J.H. Rose, Phys. Rev. B, 35 (1987) 1945-1953.
[10] F.D. Stacey, and P.M. Davis, Phys. Earth Planet. Int, 142 (2004) 137-184.
[11] R.J. Angel, Reviews in Mineralogy and Geochemistry 41 (2000) 35-60.
[12] J. Garai, CALPHAD, 30 (2006) 354-356.
[13] J. Garai, J. Applied Phys., 102 (2007) 123506.
[14] J. Garai, and A. Laugier, J. Applied Phys., 101 (2007) 023514.
[15] H. Akaike, Proc. Second Int. Symp. on Information Theory, 1973, pp. 267-281.
[16] H. Akaike, IEEE Transactions on Automatic Control 19 (1974) 716-723.
[17] P.W. Bridgman, Proc. Am. Arts. Sci., 76 (1948) 55-70.
[18] K.K. Mao, and P.M. Bell, J. Geophys. Res., 84 (1979) 4533-4536.
[19] C. Meade, and R. Jeanloz, J. Geophys. Res., 93 (1988) 3261-3269.
[20] D.J. Weidner, Y. Wang, and M.T. Vaughan, Geophys. Res. Lett., 21 (1994) 753–756.
[21] T.S. Duffy, R.J. Hemley, and H-K. Mao, H-K., Phys. Rev. Lett., 74 (1995) 1371-1374.
[22] G. Fiquet, D. Andrault, J-P. Itie, P. Gillet, and P. Richet, Phys. Earth Planet. Inter., 95 (1996) 1-17.
[23] W. Utsumi, D.J. Weidner, and R.C. Liebermann, Properties of Earth and Planetary Materials at High Pressure and Temperature, Geophys. Monogr. Ser. 101 (1998) 327-333.

[24] G. Chen, R.C. Liebermann, and D.J. Weidner, Science, 280 (1998) 1913 – 1916.
[25] Y. Fei, Amer. Mineralogist, 84 (1999) 272-276.
[26] G. Fiquet, P. Richet, and G. Montagnac, Phys. Chem. Minerals., 27 (1999) 103-111.
[27] A. Dewaele, and G. Fiquet, J. Geophys. Res., 105 (2000) 2869-2877.
[28] J. Zhang, Phys. Chem. Minerals, 27 (2000) 145-148.
[29] C-S. Zha, H-K. Mao, and R.J. Hemley, Proc. Nat. Acad. Sci., 97 (2000) 13494-13499.
[30] S. Speziale, C-S. Zha, T.S. Duffy, R.J. Hemley, and H-K. Mao, J. Geophys. Res., 106 (2001) 515-528.
[31] S. Merkel, H.R. Wenk, J. Shu, G. Shen, P. Gillet, H-K. Mao, and R.J. Hemley, J. Geophys. Res., 107 (2002) 2271.
[32] K. Hirose, N. Sata, T. Komabayashi, and Y. Ohishi, Phys. Earth Planet. Int. 167 (2008) 149–154.
[33] C-S. Zha, K. Mibe, W.A. Basset, O. Tschauner, H-K. Mao, and R.J. Hemley, J. Appl. Phys., 103 (2008) 054908.
[34] D. G. Isaak, R.E. Cohen, and M.J. Mehl, J. Geophys. Res., 95 (1990) 7055-7067.
[35] I. Inbar, and R.E. Cohen, Geophys. Res. Lett., 22 (1995) 1533-1536.
[36] J. Hama, and K. Suito, Phys. Earth Planet. Int., 114 (1999) 165-179.
[37] M.H.G. Jacobs, and H.A.J. Oonk, Phys. Chem. Chem. Phys., 2 (2000) 2641-2646.
[38] M.H.G. Jacobs, and H.A.J. Oonk, CALPHAD, 24 (2000) 133-148.
[39] B.B. Karki, R.M. Wentzcovitch, S. De Gironcoli, and S. Baroni, Phys. Rev. B, 61 (2000) 8793-8800.
[40] Y. Fei, J. Li, K. Hirose, W. Minarik, J. Van Orman, C. Sanloup, W. van Westrenen, T. Komabayashi, and K. Funakoshi, Earth Planet. Int., 143-144 (2004) 515-526.
[41] Y. Zhang, D. Zhao, M. Matsui, and G. Guo, J. Geoph. Res., 112 (2007) B11202.
[42] Z. Wu, R.M. Wentzcovitch, K. Umemoto, N. Li, K. Hirose, and J. Zheng, J. Geoph. Res., 113 (2008) B06204.
[43] L.S. Dubrovinsky, and S.K. Saxena, Phys. Chem. Minerals, 24 (1997) 547-550.
[44] S.V. Sinogeikin, J.M. Jackson, B. O'Neill, J.W. Palko and J.D. Bass, Rev. Sci. Instrum., 71 (2000) 201-206.
[45] Y. Sumino, O.L. Anderson and I. Suzuki, Phys. Chem. Minerals, 9 (1983) 38-47.
[46] D.G. Isaak, O.L. Anderson and T. Goto, Phys. Chem. Minerals, 16 (1989) 704-713.
[47] P. Debye, Ann. Phys. 39 (1912) 789.
[48] Landolt-Bornstein, Zahlenwerte und Funktionen aus Physic, Chemie, Astronomie, Geophysic, und Technik, II. Band, Eigenschaften der Materie in Ihren Aggregatzustanden, 4 Teil, Kalorische Zustandsgrossen, Springer-Verlag, 1961.
[49] I.S. Grigoriev, E.Z. Meilikhov, Handbook of Physical Quantities, CRC Press, Inc. Boca Raton, FL, USA, 1997.
[50] I. Barin, Thermochemical Data of Pure Substances, VCH, Weinheim, Federal republic of Germany, 1989.

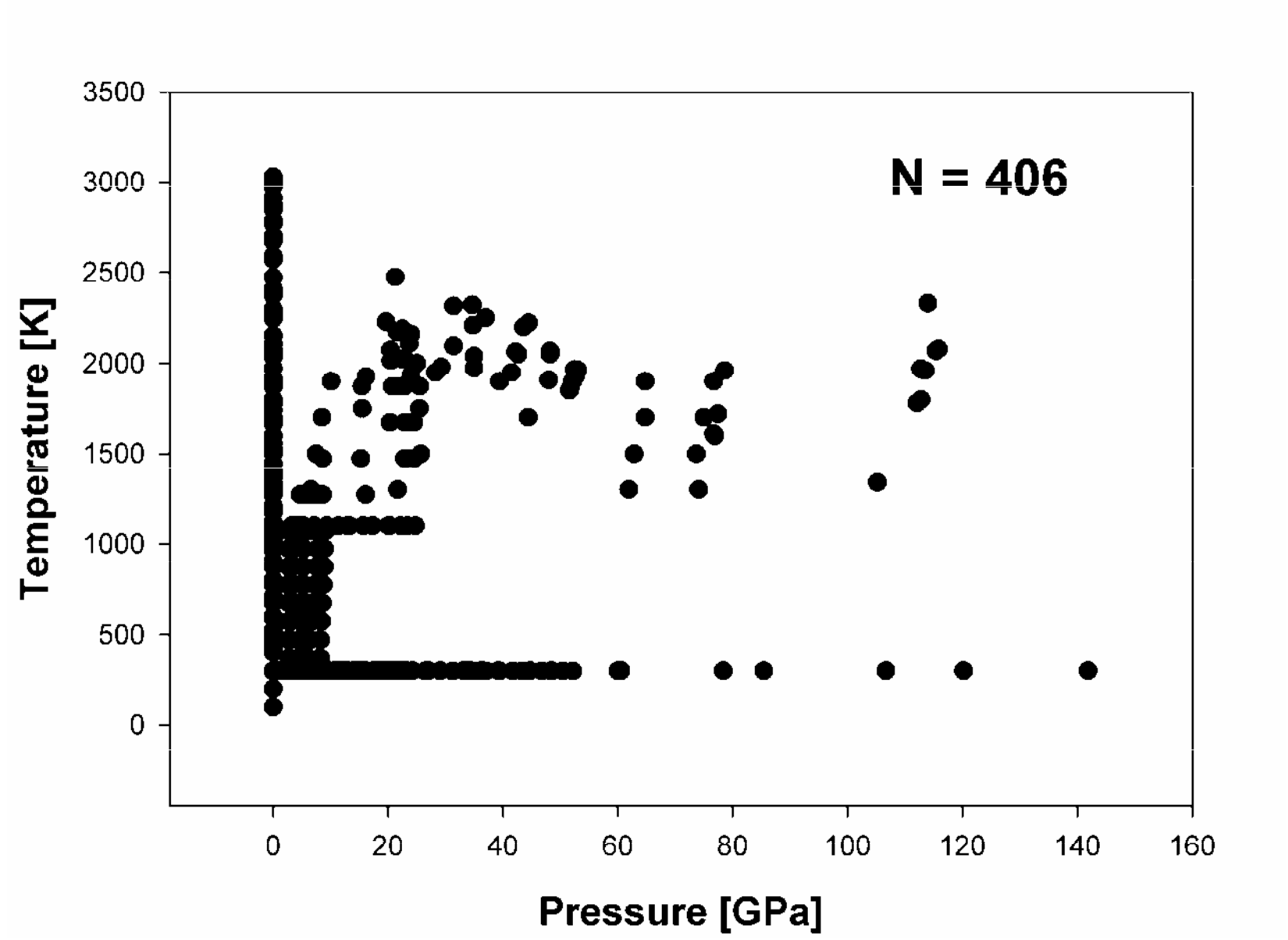

**Fig. 1**. The pressure temperature distribution of the experimental data.

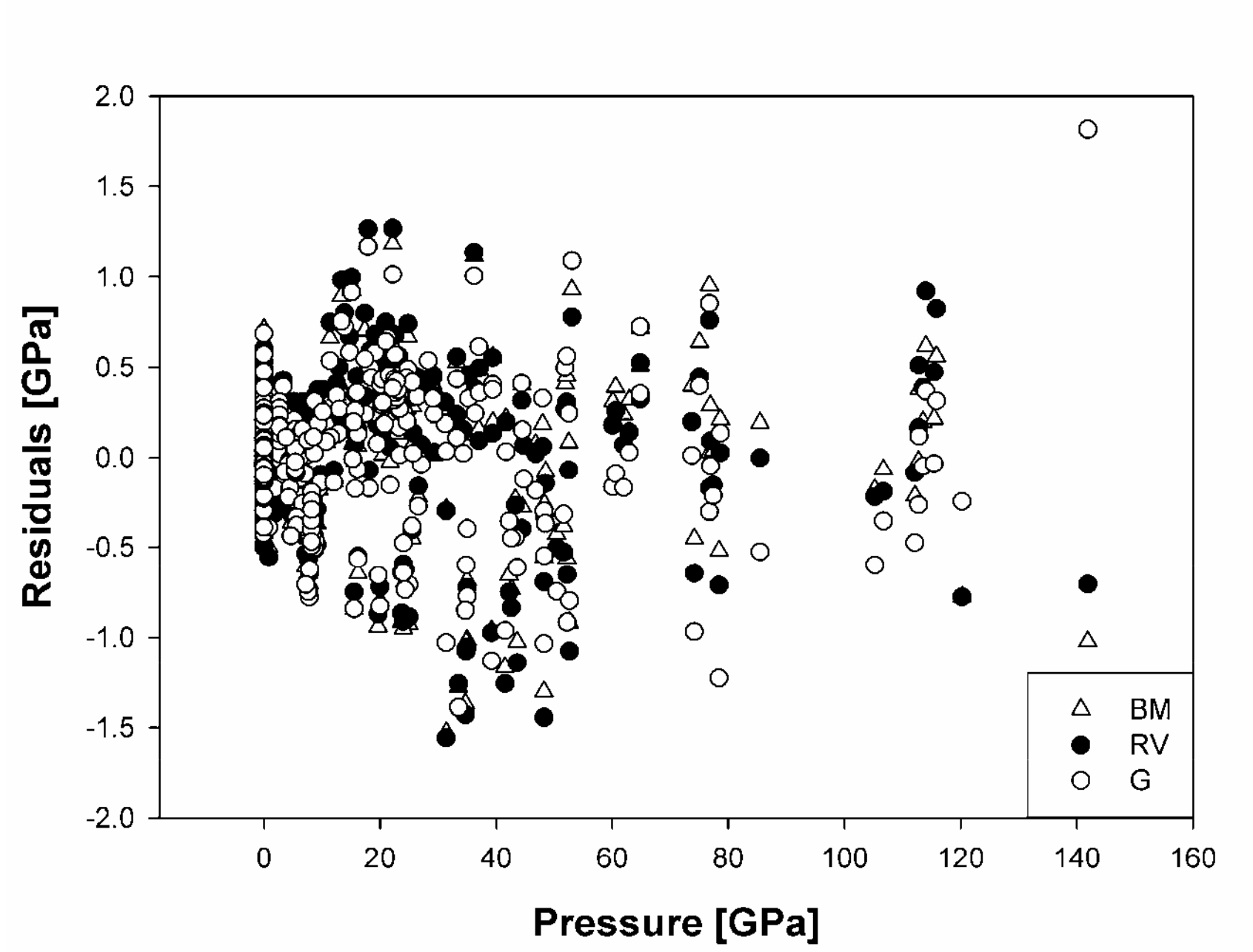

**Fig. 2**. Residuals of the three EoSs are plotted against the pressure.

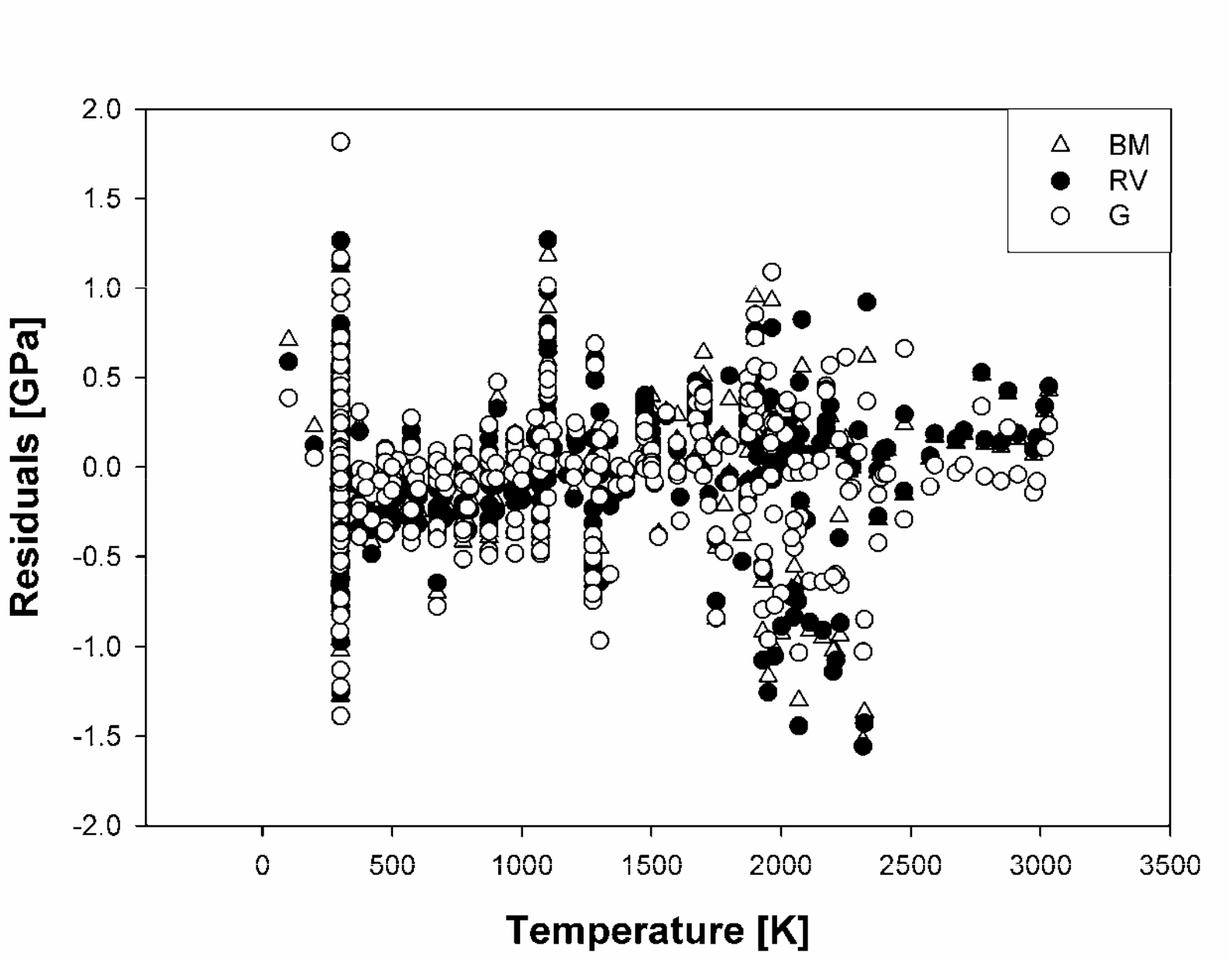

**Fig. 3.** Residuals of the three EoSs are plotted against the temperature.

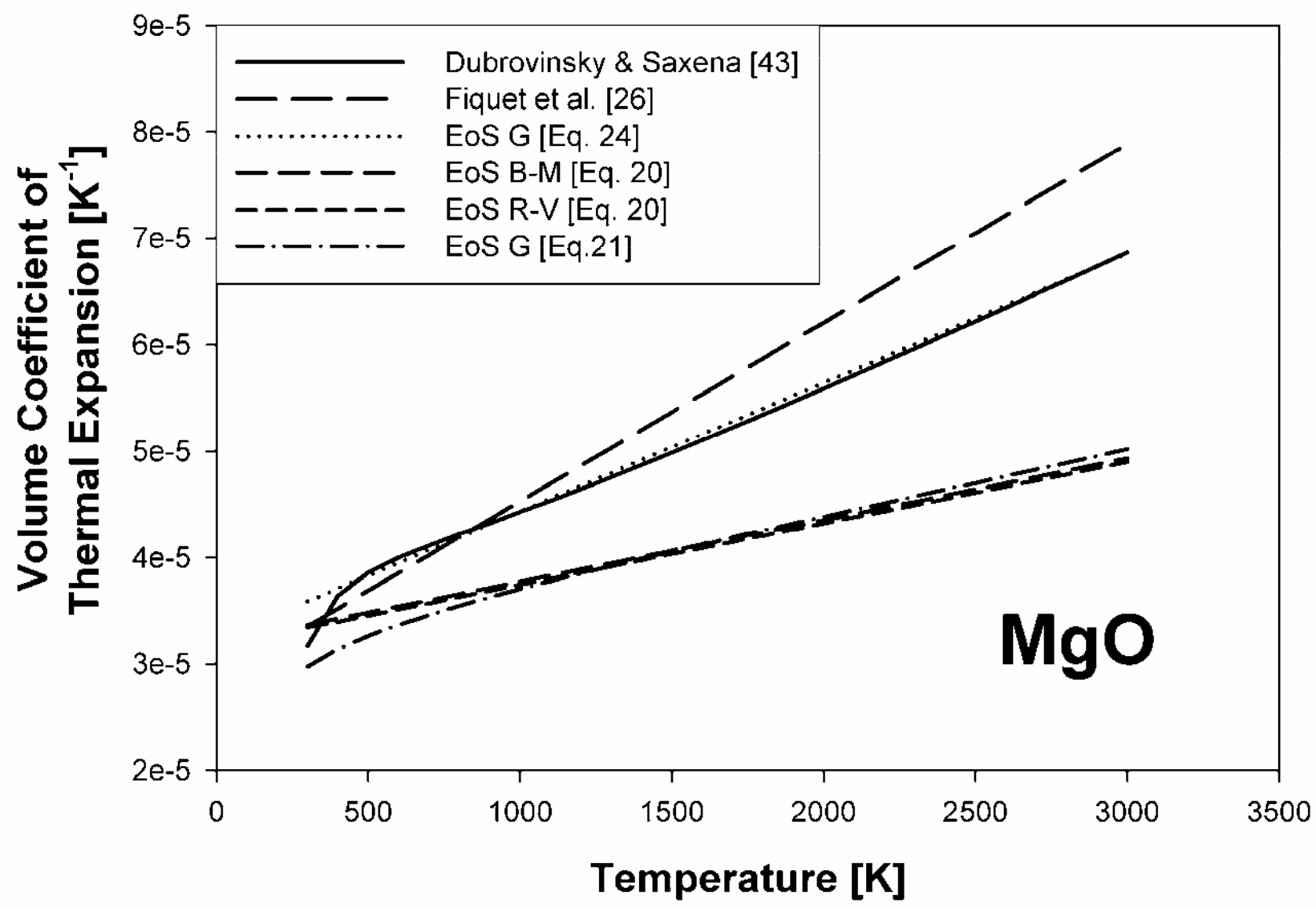

**Fig. 4**. Volume coefficient of thermal expansion values are plotted against the temperature. The difference between the experimental values and the values calculated from the EoS are explained in the text. The "instantaneous" volume coefficient of thermal expansion values calculated by Eq. (24) fit well to the experiments.

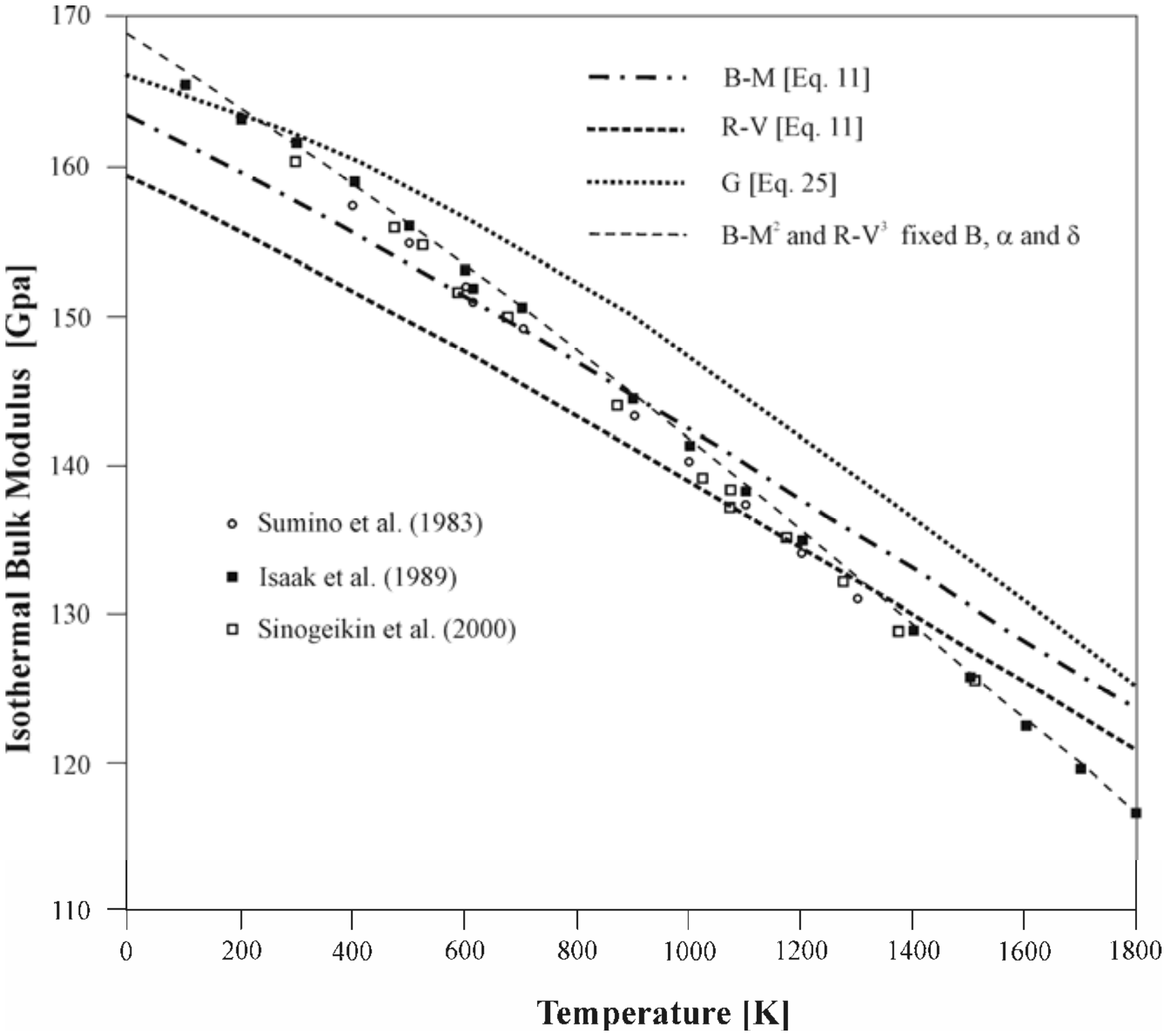

**Fig. 5.** Calculated isothermal bulk modulus values are plotted against experimental data. The thermodynamic parameters for the EoSs are given in Table 1.

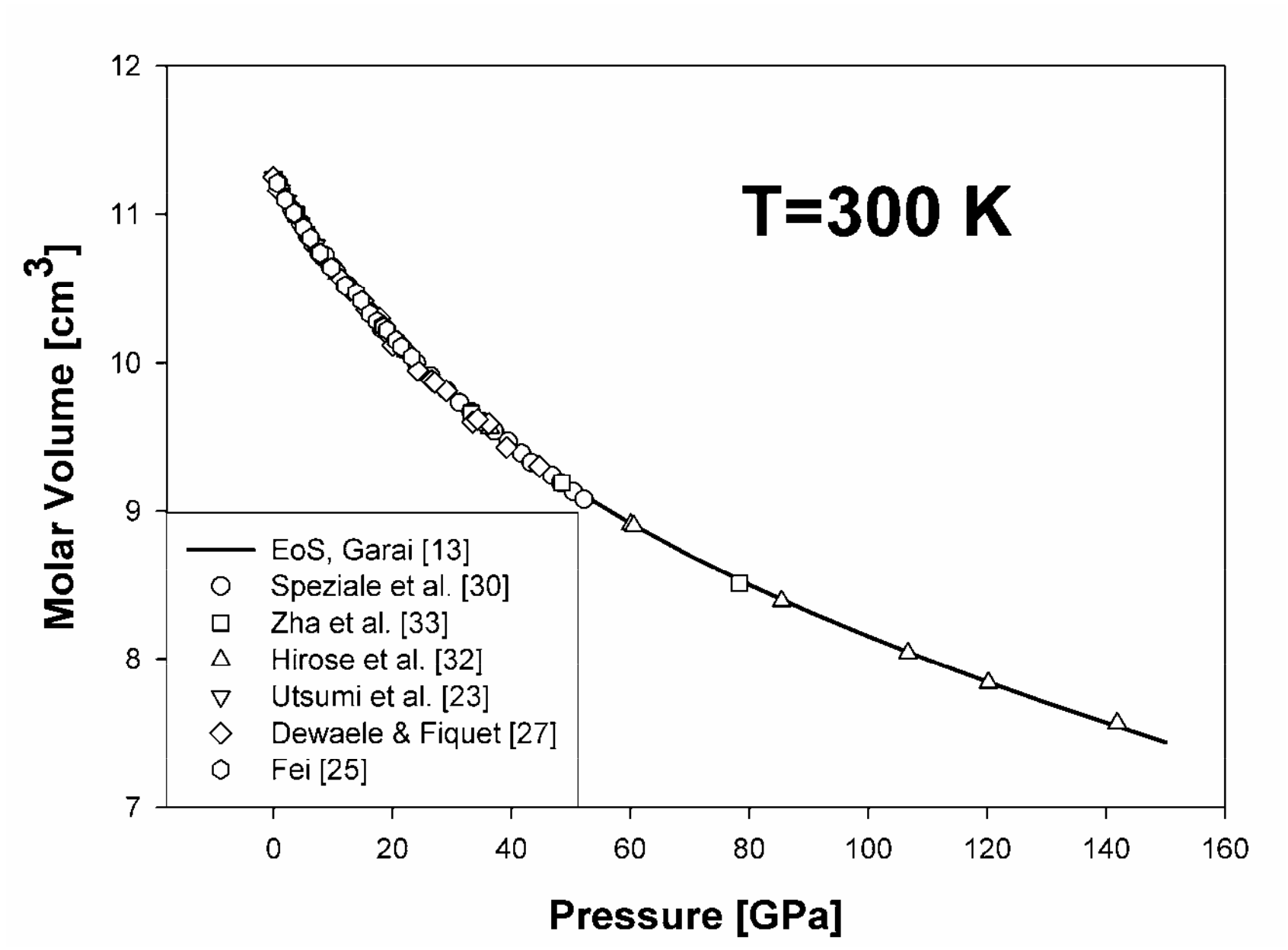

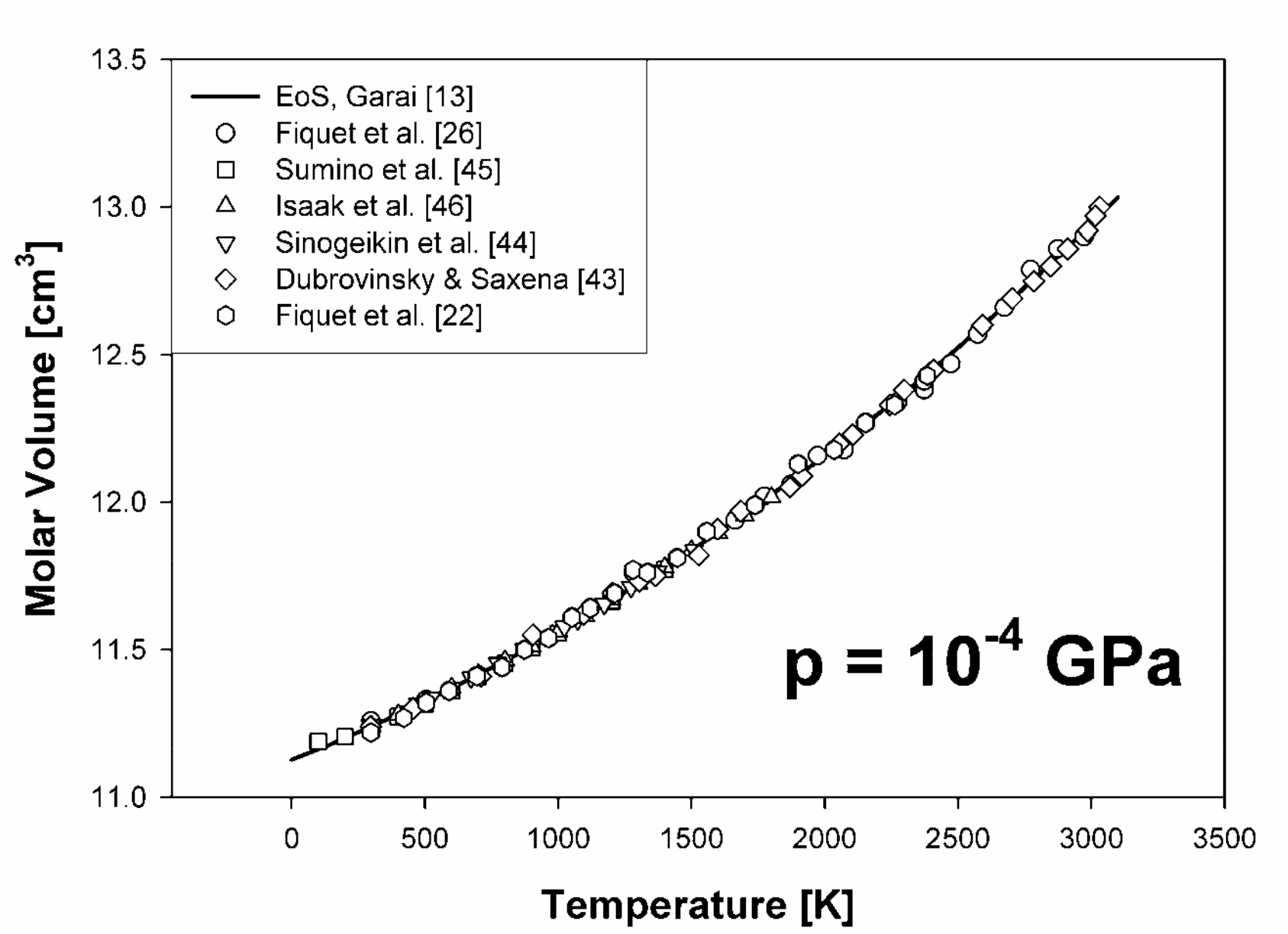

**Fig. 6.** Experiments are plotted against the EoS of G at ambient condition. (a) pressure-volume (b) temperature-volume

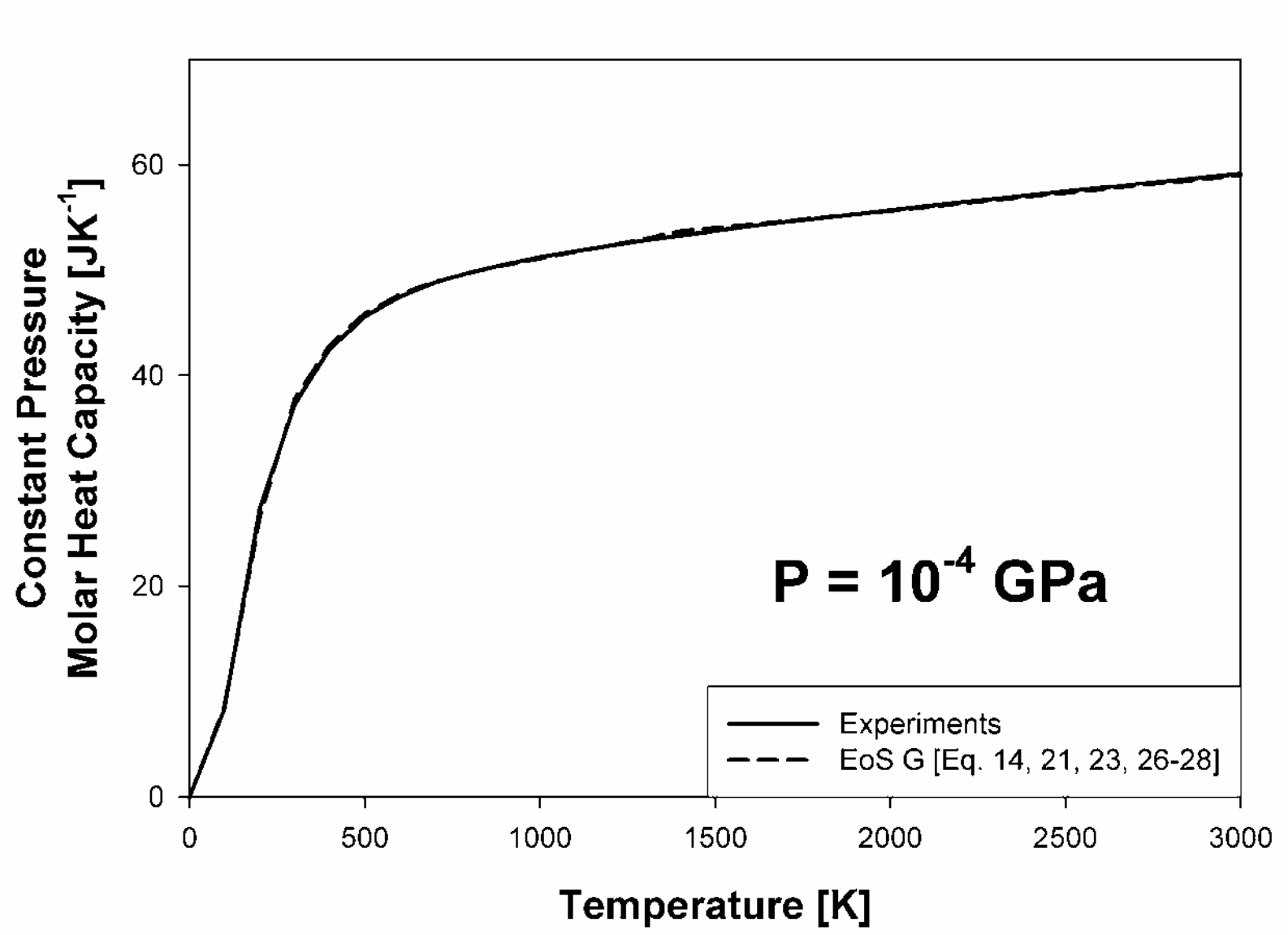

**Fig. 7.** Calculated constant pressure molar heat capacity is plotted against experiments at 1 bar pressure.

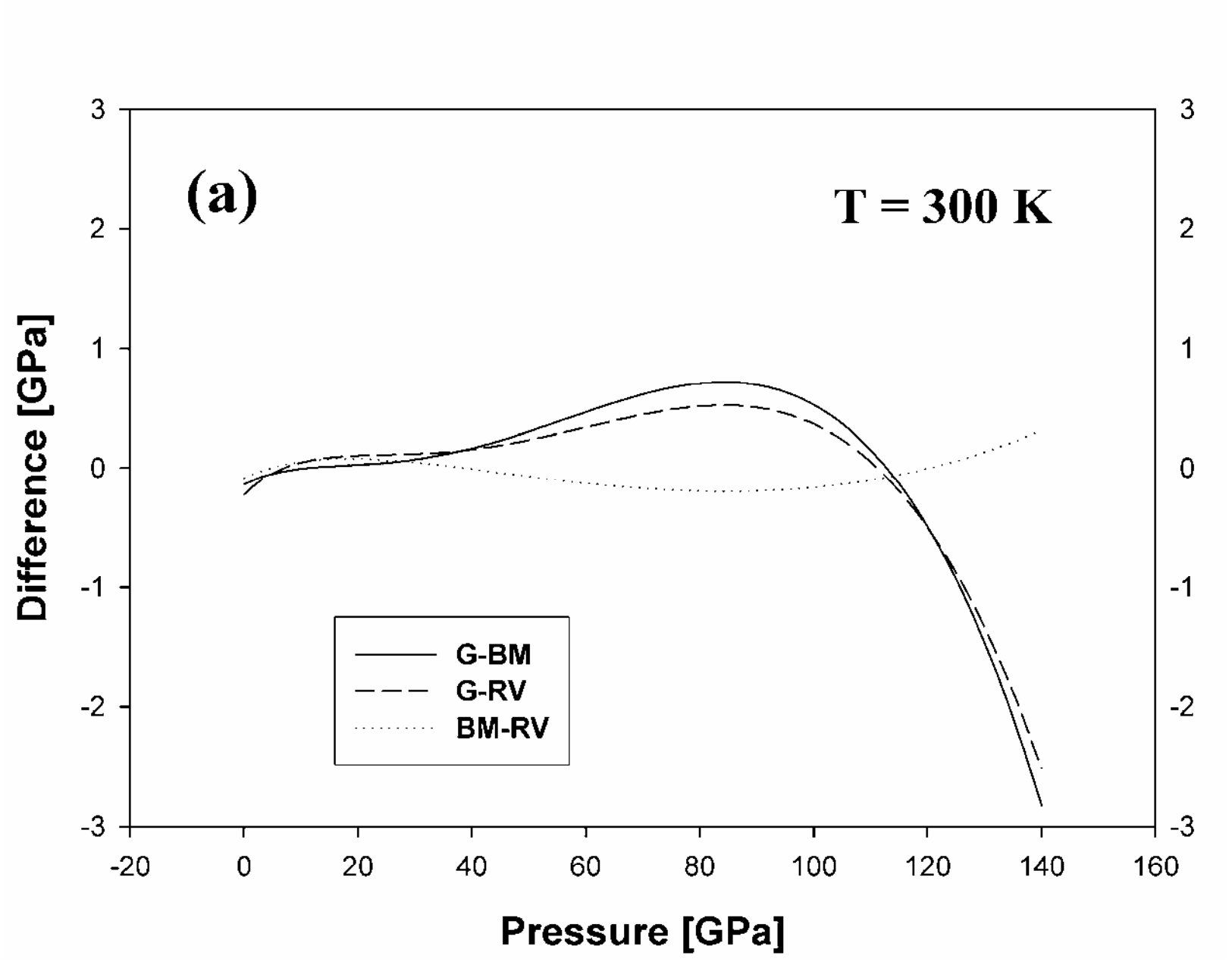

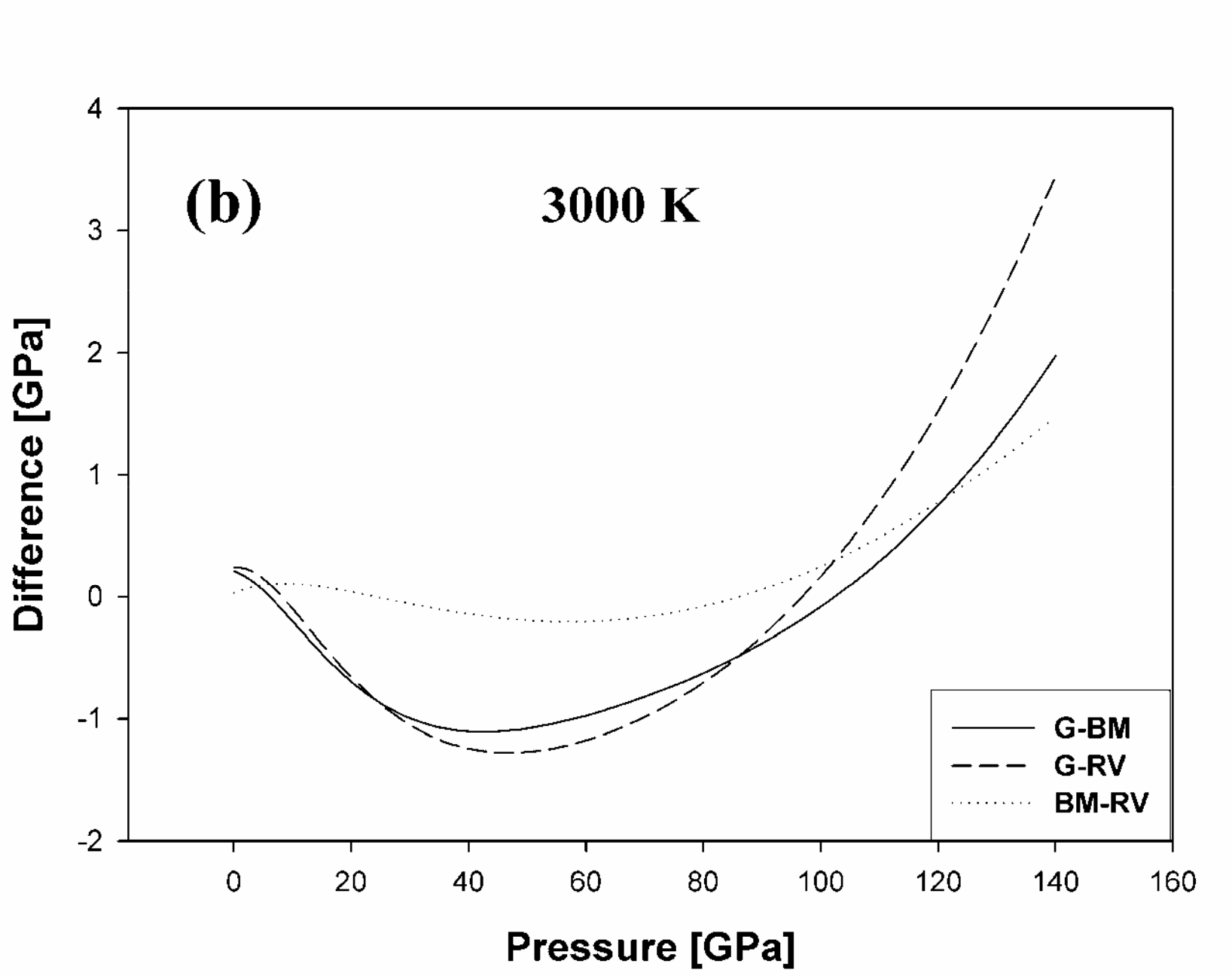

**Fig. 8**. Comparison of the three EoSs. The pressure differences are plotted at 300 and 3000 K temperatures. (a) 300 K, and (b) 3000 K.

**Table 1.** Parameters describing the p-V-T relationship of periclase (MgO).

| EoS | | $V_o$ [$cm^3$] | $K_o$ [GPa] | $K_o'$ | $\alpha_o$ $[\times 10^{-5} K^{-1}]$ | $\alpha_1$ $[\times 10^{-9} K^{-2}]$ | $\delta$ |
|---|---|---|---|---|---|---|---|
| B-M | (p-V-T) [Eq. 12] | 11.137 | 163.59 | 4.145 | 3.197 | 5.784 | 3.690 |
| | | 11.109 | 168.35*1 | 4.072 | 3.529 | 4.684 | 3.758 |
| | | 11.125 | 168.83*2 | 4.292 | 2.950*2 | 8.354 | 4.60*2 |
| R-V | (p-V-T) [Eq. 13] | 11.145 | 159.42 | 4.499 | 3.169 | 5.767 | 3.667 |
| | | 11.140 | 168.66*1 | 4.330 | 3.691 | 4.025 | 3.769 |
| | | 11.118 | 168.91*3 | 4.569 | 2.960*3 | 8.226 | 4.65*3 |

| EoS | $V_o$ [$cm^3$] | $K_o$ [GPa] | $K_{1P}$ | $K_{2P}$ $[\times 10^{-3} GPa^{-1}]$ | $\alpha_o$ $[\times 10^{-5} K^{-1}]$ | $\alpha_{1P}$ $[\times 10^{-7} GPa^{-1} K^{-1}]$ | $\alpha_{2P}$ $[\times 10^{-9} GPa^{-2} K^{-1}]$ | $\alpha_{1T}$ $[\times 10^{-9} K^{-2}]$ | a |
|---|---|---|---|---|---|---|---|---|---|
| G [Eq.14] | 11.127 | 165.98 | 1.843 | -3.233 | 3.227 | -3.511 | 1.7 | 6.043 | 6.28 |

*fixed [$K_{0,\, T=298\, K}$ = 161.6 GPa, and $\alpha_{0,T=298K} = 3.20 \times 10^{-5} K^{-1}$]

B-M = Birch-Murnaghan EoS; R-V = Rydberg-Vinet EoS; G = Garai EoS

**Table 2.** Fitting parameters for the EoSs.

| Pressure Range [GPa] | Temperature Range [K] | Number of Experiment | EoS | Volume | | Pressure | | Temperature | |
|---|---|---|---|---|---|---|---|---|---|
| | | | | RMS Misfit [$cm^3$] | AIC | RMS Misfit [GPa] | AIC | RMS Misfit [K] | AIC |
| 0-141.8 | 80-3031 | 406 | B-M | - | - | 0.381 | -770.7 | - | - |
| | | | | - | - | 0.469*1 | -604.3*1 | - | - |
| | | | | - | - | 0.921*2 | -60.9*2 | - | - |
| | | | R-V | - | - | 0.396 | -741.1 | - | - |
| | | | | - | - | 0.466*1 | -609.4*1 | - | - |
| | | | | - | - | 1.005*3 | 9.7*3 | - | - |
| | | | G | 0.018 | -3239.9 | 0.371 | -804.5 | 60.3 | 3328.6 |

*Fixed parameters

[1][$K_{0,T=298\ K}$ = 161.6 GPa]

[2][$K_{0,T=298\ K}$ = 161.6 GPa, $\alpha_{0,T=298K} = 3.20\times10^{-5}\ K^{-1}$ and $\delta = 4.60$]

[3][$K_{0,T=298\ K}$ = 161.6 GPa, $\alpha_{0,T=298K} = 3.20\times10^{-5}\ K^{-1}$ and $\delta = 4.65$]

**Table 3.** Fitting parameters for the individual investigations.

| Experiment | Pressure Range [GPa] | Temperature Range [K] | Number of Experiment | EoS | Volume | | Pressure | | Temperature | |
|---|---|---|---|---|---|---|---|---|---|---|
| | | | | | RMS Misfit [$cm^3$] | AIC | RMS Misfit [GPa] | AIC | RMS Misfit [K] | AIC |
| Fiquet et al. 1996 | $10^{-4}$ | 298-2385 | 21 | B-M | - | - | 0.260 | -56.6 | - | - |
| | | | | R-V | - | - | 0.271 | -54.9 | - | - |
| | | | | G | 0.0189 | -166.7 | 0.221 | -63.4 | 33.6 | 147.6 |
| Dubrovinsky & Saxena, 1997 | $10^{-4}$ | 298-3031 | 27 | B-M | - | - | 0.197 | -87.7 | - | - |
| | | | | R-V | - | - | 0.209 | -84.7 | - | - |
| | | | | G | 0.0140 | -230.3 | 0.153 | -101.4 | 23.0 | 169.3 |
| Utsumi et al. 1998 | 0-9.52 | 300-1273 | 61 | B-M | - | - | 0.280 | -155.2 | - | - |
| | | | | R-V | - | - | 0.248 | -170.0 | - | - |
| | | | | G | 0.0203 | -475.6 | 0.324 | -137.6 | 51.0 | 479.7 |
| Fei, 1999 | 0-24.8 | 300, 1100 | 35 | B-M | - | - | 0.467 | -53.4 | - | - |
| | | | | R-V | - | - | 0.521 | -45.6 | - | - |
| | | | | G | 0.0203 | -272.8 | 0.404 | -63.5 | 63.8 | 290.9 |
| Fiquet et al. 1999 | $10^{-4}$ | 298-2973 | 36 | B-M | - | - | 0.220 | -108.2 | - | - |
| | | | | R-V | - | - | 0.227 | -106.6 | - | - |
| | | | | G | 0.0217 | -275.7 | 0.221 | -108.7 | 33.5 | 252.9 |
| Dewaele et al., 2000 | 0-53.0 | 300-2474 | 54 | B-M | - | - | 0.728 | -34.2 | - | - |
| | | | | R-V | - | - | 0.751 | -30.9 | - | - |
| | | | | G | 0.0262 | -393.4 | 0.633 | -49.4 | 103.9 | 501.5 |
| Zhang, 2000 | 0-8.2 | 300-1073 | 30 | B-M | - | - | 0.192 | -99.1 | - | - |
| | | | | R-V | - | - | 0.161 | -109.5 | - | - |
| | | | | G | 0.0145 | -254.1 | 0.231 | -88.0 | 37.0 | 216.6 |
| Speziale et al., 2001 | 0-52.2 | 298 | 32 | B-M | - | - | 0.330 | -71.0 | - | - |
| | | | | R-V | - | - | 0.357 | -65.9 | - | - |
| | | | | G | 0.0135 | -275.4 | 0.359 | -65.5 | 56.6 | 258.4 |
| Fei et al., 2004 | 8.6-25.6 | 1273-2173 | 26 | B-M | - | - | 0.235 | -75.3 | - | - |
| | | | | R-V | - | - | 0.300 | -62.5 | - | - |
| | | | | G | 0.0161 | -214.8 | 0.316 | -60.0 | 51.3 | 204.8 |
| Hirose et al., 2007 | 10.7-141.8 | 300-2330 | 21 | B-M | - | - | 0.401 | -38.3 | - | - |
| | | | | R-V | - | - | 0.433 | -35.1 | - | - |
| | | | | G | 0.0070 | -208.5 | 0.413 | -37.1 | 70.2 | 178.6 |
| Zha et al., 2008 | 8.5-11.27 | 300-1900 | 20 | B-M | | | 0.409 | -35.8 | | |
| | | | | R-V | | | 0.370 | -39.8 | | |
| | | | | G | 0.0121 | -176.5 | 0.480 | -29.4 | 82.8 | 176.7 |
| Sinogeikin et al., 2000 | $10^{-4}$ | 295-1510 | 15 | B-M | | | 0.047 | -91.6 | | |
| | | | | R-V | | | 0.099 | -69.4 | | |
| | | | | G | 0.0041 | -165.0 | 0.053 | -88.0 | 8.5 | 64.2 |
| Isaak et al., 1989 | $10^{-4}$ | 300-1800 | 16 | B-M | | | 0.073 | -83.9 | | |
| | | | | R-V | | | 0.122 | -67.4 | | |
| | | | | G | 0.0028 | -188.5 | 0.030 | -111.8 | 4.6 | 48.7 |
| Sumino et al., 1983 | $10^{-4}$ | 80-1300 | 12 | B-M | | | 0.273 | -31.2 | | |
| | | | | R-V | | | 0.284 | -30.2 | | |
| | | | | G | 0.0101 | -110.2 | 0.143 | -46.6 | 24.7 | 77.0 |

B-M = Birch-Murnaghan EoS; V = Vinet EoS; G = Garai, 2007